# Effect of Mixed Precision Computing on H-Matrix Vector Multiplication in BEM Analysis

Rise Ooi
Hokkaido University
rise.ooi@frontier.hokudai.ac.jp

Takeshi Iwashita
Hokkaido University
iwashita@iic.hokudai.ac.jp

Takeshi Fukaya
Hokkaido University
fukaya@iic.hokudai.ac.jp

Akihiro Ida
The University of Tokyo
ida@cc.u-tokyo.ac.jp

Rio Yokota
Tokyo Institute of Technology
rioyokota@gsic.titech.ac.jp

## ABSTRACT

Hierarchical Matrix (H-matrix) is an approximation technique which splits a target dense matrix into multiple submatrices, and where a selected portion of submatrices are low-rank approximated. The technique substantially reduces both time and space complexity of dense matrix vector multiplication, and hence has been applied to numerous practical problems.

In this paper, we aim to accelerate the H-matrix vector multiplication by introducing mixed precision computing, where we employ both binary64 (FP64) and binary32 (FP32) arithmetic operations. We propose three methods to introduce mixed precision computing to H-matrix vector multiplication, and then evaluate them in a boundary element method (BEM) analysis. The numerical tests examine the effects of mixed precision computing, particularly on the required simulation time and rate of convergence of the iterative (BiCG-STAB) linear solver. We confirm the effectiveness of the proposed methods.

## 1 INTRODUCTION

*Hierarchical Matrix* (H-matrix) [11] is an approximation technique for dense matrices. The technique approximates a target dense matrix by splitting it up into multiple submatrices, where the selected submatrices are low-rank approximated. For the low-rank approximation of the submatrices, methods such as random sampling, *Singular Value Decomposition* (SVD), and *Adaptive Cross Approximation* (ACA) [18] are well known. In an ideal case, the technique reduces both time and space complexity of $O(N^2)$ of (dense) matrix vector multiplication (matvecmul) into $O(N \log N)$, where $N$ is the dimension of the matrix, assuming it is square. Therefore, H-matrices have been effectively used for practical problems, including $n$-body [25, 28], earthquake cycles [22], and superconductive coils [24] simulations. In this paper, we aim to accelerate the H-matrix vector multiplication by introducing mixed precision computing, where we use both binary64 (FP64) and binary32 (FP32) [1] arithmetic operations.

In recent years, mixed (and lower) precision computing has been investigated in various computational kernels and scientific applications. Under the demands for more computing needs, there exist modern computational devices that can process low precision computations much faster than higher ones, for example, in some types of graphics processing units (GPU). Notably, we see trends of deep learning-based applications perform sufficiently well at lower accuracy computations, especially when they are computed using binary16 (FP16) and utilizing Nvidia's Tensor Cores [21]. Moreover, in the case of memory-bound applications such as matrix vector multiplication, the representation of data in lower precision simply reduces the amount of data transferred between CPU and memory, which naturally results in better performance.

In this paper, we propose three methods to introduce mixed precision computing to H-matrix vector multiplication, and evaluate them in a boundary element method (BEM) analysis. BEM [4] is one of the most popular discretization methods for partial differential equation problems, and is also one of the principal application domains of H-matrix and its related techniques. In a BEM analysis, a linear system of equations with a dense coefficient matrix is to be solved. If a Krylov subspace iterative method [23] is used for the solution process, then the intensely iterated coefficient matrix vector multiplication is the most time consuming constraint part. Here, the H-matrix technique efficiently approximates the dense coefficient matrix and thus significantly accelerates the matrix vector multiplication operations. Although when we introduce the mixed precision data representation for the matrix, it usually accelerates the matrix vector multiplication operation, this approach comes with a side effect, that is the decline of accuracy of operations. In the case of a Krylov subspace solver, the convergence rate generally degrades. And so accordingly, we conduct a boundary element electrostatic field analysis to examine both the positive and negative effects of mixed precision computing in H-matrix vector multiplication. Our numerical tests confirm the effectiveness of the proposed mixed precision computing methods in the test analysis. This newly developed mixed precision technique is subsequently being used to enhance the open source H-matrix library HACApK [15–17]. We choose the HACApK libary for its advantages in its implementation based on hybrid parallel programming model which utilizes both multithreading and multiprocessing, its support for general x86 Linux clusters and K computer, and its active development and usage in various applications.

## 1.1 Related works and our main contribution

Mixed precision computing has been studied for a long time mainly in the context of iterative refinement for the solution of a linear system [7, 14], where extra precision (e.g. double-double precision) is partially employed to obtain more accurate solutions to certain ill-conditioned problems. In the 2000s, since FP32 is generally about twice faster than FP64 on a standard processor, the LU factorization based iterative refinement using FP32 and FP64 was actively studied [5, 19]. In addition, FP32 was much faster than FP64 on early generation GPUs, and it motivated researchers to investigate the mixed precision iterative (Krylov) solvers using FP32 and FP64 [3]. Recently, due to the demand in AI applications, FP16 has become strongly supported on GPUs (and some CPUs). In order to exploit the high potential of FP16, extensions of the traditional iterative refinement have been studied, where mixed precision computing using multiple precision, i.e. FP64, FP32, and FP16, are proposed and analyzed [6, 12].

Recently, the implementation of H-matrix vector multiplication using Batched BLAS has been reported [27], and Batched BLAS kernels using FP16 have also been presented [2]. Combining these results may realize the mixed precision H-matrix vector multiplication, but in contrast to linear solvers, and to the best of our knowledge, there is currently no report on mixed precision computing for H-matrix, especially on H-matrix vector multiplication. Consequently, the main contribution of this paper is the proposal of the three methods to introduce mixed precision computing to H-matrix and the further enhancement of the H-matrix library with mixed precision computing.

## 2 HIERARCHICAL MATRICES

A Hierarchical Matrices (H-Matrix) [11] approximates a dense matrix so that it has a much better space order of complexity than its original dense matrix. Although a dense square matrix requires an order of $O(N^2)$, the H-Matrix version requires an order of about $O(N \log N)$ in its ideal case, excluding all other constant parameters.

Let $A$ be an $N \times N$ real dense matrix, we denote each of its submatrices as $A|^m$, where $m$ is the identifier. We define $I_m$ and $J_m$ as the sets of row and column indexes (in the original matrix) of a submatrix $A|^m$ respectively, and denote the first and last indexes by $i_m^{(s)}(j_m^{(s)})$ and $i_m^{(e)}(j_m^{(e)})$ respectively. Then, we let $M$ be a set of $m$.

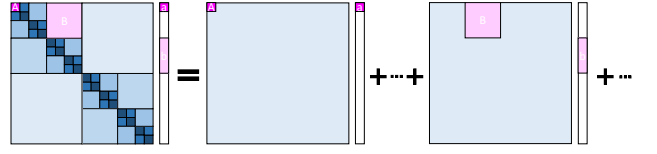

Figure 1: An illustration of H-matrix vector multiplication.

Now, we consider the case that $M$ satisfies

$$(I_{m_1} \times J_{m_1}) \cap (I_{m_2} \times J_{m_2}) = \emptyset \quad (\forall m_1, m_2 \in M) \quad (1)$$

and

$$\bigcup_{m \in M} (I_m \times J_m) = I \times J, \quad (2)$$

where $I$ and $J$ are the sets of row and column indexes of the original matrix $A$ respectively. In this case, $M$ represents a way of partitioning $A$ into a set of submatrices.

A H-matrix approximation of $A$ is based on the submatrix partitioning, where for the target matrix $A$, we must first find a way to partition $M$. Then, for each submatrix $A|^m$ ($m \in M$), we determine whether if we can apply low-rank approximation to $A|^m$. Here, the low-rank approximation means

$$A|^m \approx V_m W_m, \quad (3)$$

where $V_m \in \mathbb{R}^{\#I_m \times r_m}$, $W_m \in \mathbb{R}^{r_m \times \#J_m}$, $r_m \leq \min(\#I_m, \#J_m)$. Note that $\#I_m$ and $\#J_m$ are the number of members (elements) of $I_m$ and $J_m$ respectively. Let $M^{(LR)} \subseteq M$ be a set of $m$, where the low-rank approximation is applicable to $A|^m$, and the resulting H-matrix approximation of $A$, which we denote by $\widetilde{A}$, is obtained as

$$\widetilde{A}|^m := \begin{cases} V_m W_m & (m \in M^{(LR)}) \\ A|^m & (m \notin M^{(LR)}) \end{cases} \quad (m \in M). \quad (4)$$

### 2.1 H-matrix vector multiplication

We consider the H-matrix vector multiplication to be

$$\widetilde{A} x \to \widetilde{y}, \quad (5)$$

where $x, \widetilde{y} \in \mathbb{R}^N$ are the source and result vectors respectively. Since a H-matrix essentially consists of just a set of submatrices, the H-matrix vector multiplication can then be computed by a set of operations for each submatrix. More precisely, for a submatrix $\widetilde{A}|^m (\in \mathbb{R}^{\#I_m \times \#J_m})$, consider that

$$\widetilde{A}|^m x_m \to \hat{y}_m \in \mathbb{R}^{\#I_m}, \quad (6)$$

where $x_m := (x[j_m^{(s)}], x[j_m^{(s)}+1], \ldots, x[j_m^{(e)}])^\top \in \mathbb{R}^{\#J_m}$ and $x[i]$ means the $i$-th element of the vector $x$. Then, we can define $\bar{y}_m \in \mathbb{R}^N$ as

$$\bar{y}_m[i] := \begin{cases} 0 & (i \notin I_m) \\ \hat{y}_m[i - i_m^{(s)} + 1] & (i \in I_m) \end{cases} \quad (i = 1, 2, \ldots, N), \quad (7)$$

and finally obtain

$$\widetilde{y} = \sum_{m \in M} \bar{y}_m, \quad (8)$$

which is illustrated in Figure 1. Concretely, in the case of submatrix $\widetilde{A}|^m$ being dense, Equation (6) is computed just like the standard matrix vector multiplication; on the other hand, if a submatrix is

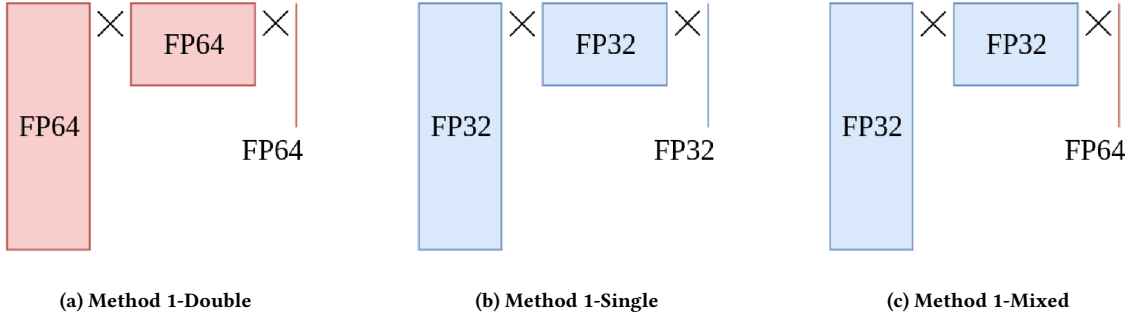

Figure 2: An illustration of the three different low-rank matrix vector multiplication models for Method 1.

low-rank, then Equation (6) is computed through the following two steps:

$$W_m\, x_m \to z_m \in R^{r_m}, \quad (9)$$

and

$$V_m\, z_m \to \hat{y}_m. \quad (10)$$

## 3 METHODOLOGY

In this paper, we propose three methods to introduce the mixed precision computing into H-Matrix vector multiplication. In each method, we assume the data structure of a H-matrix (and source vector) and determine each data type to be in either FP64 or FP32 precision. The overview of our proposed methods are as follows:

- **Method 1**: We use the original data structure of a H-matrix multiplied by a source vector, but assign various combinations of data types to them.
- **Method 2**: We modify (further expanding) the data structure that represents the low-rank submatrices of the H-matrix in attempt to alleviate the drawback of using lower precision. Again, the H-matrix and source vector are assigned a variety of precision.
- **Method 3**: We further modify the data structure proposed in Method 2 by introducing more sophisticated combinations of precision types.

For all of our proposed methods, we evaluate their performance in both H-matrix vector multiplication, and Krylov iterative solver, namely BiCGSTAB, which appears in a BEM analysis described later.

### 3.1 Method 1: Mixed precision computing using the original data structure

For our first method, we simply use the data structure employed in the conventional H-matrix library, namely HACApK, but assign various combinations of precision; we determine the precision (FP64 or FP32) for each of H-matrix (including dense and low-rank submatrices), Source Vector, and Result Vector.

In this study, we consider the following three kinds of combinations in Method 1:

- **Method 1-Double** (*Baseline* see Figure 2a):
  FP64 H-Matrix × FP64 Source Vector → FP64 Result Vector
- **Method 1-Single** (see Figure 2b):
  FP32 H-Matrix × FP32 Source Vector → FP64 Result Vector
- **Method 1-Mixed** (see Figure 2c):
  FP32 H-Matrix × FP64 Source Vector → FP64 Result Vector

Here, Source and Result Vectors correspond to $x$ and $\widetilde{y}$ in Equation (8) respectively.

*Method 1-Double* is the baseline, which uses FP64 only, this is the original version that is actively being used by the HACApK library. *Method 1-Single* let both H-matrix and Source Vector be at FP32, and store Result Vector at FP64. Finally, *Method 1-Mixed* let the H-Matrix be in FP32 but the Source Vector to be FP64, and then store Result Vector at FP64. The idea behind Method 1-Mixed is that the data storage size requirement of H-matrix is much larger than that of Source Vector. Thus, storing the H-Matrix at lower precision may accelerate the H-matrix vector multiplication computation while retaining the full precision of Source Vector may be useful. Figure 2 shows the H-Matrix's low-rank submatrix vector multiplication models.

### 3.2 Method 2: Mixed precision computing using a modified data structure

For our second method, we modify the data structure that represents low-rank submatrices by further expanding it in attempt to alleviate the drawback of using lower precision, we then assign a variety of precision precision to them. In the case of using lower precision, overflow and underflow are more likely to happen than the case of using higher precision because the exponent in lower precision is smaller than that in higher precision. In order to remedy this drawback, we modify the low-rank representation; we use

$$\widetilde{A}|^m = V'_m D_m W'_m, \quad (11)$$

instead of the original form

$$\widetilde{A}|^m = V_m W_m, \quad (12)$$

where $D_m$ is an $r_m \times r_m$ diagonal matrix.

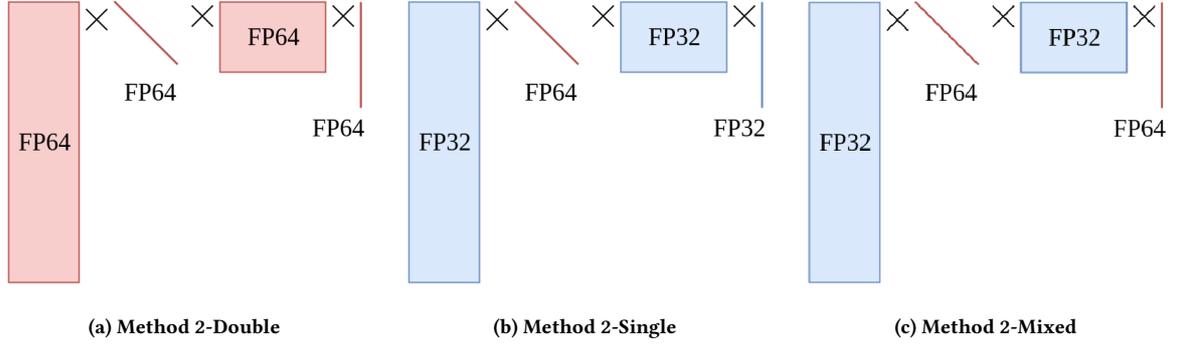

**(a) Method 2-Double**  **(b) Method 2-Single**  **(c) Method 2-Mixed**

Figure 3: An illustration of the three different low-rank matrix vector multiplication models for Method 2.

We determine the diagonal matrix $D_m$ by further decomposing $V_m$ and $W_m$. Let

$$V_m = \begin{pmatrix} v_1 & v_2 & \cdots & v_{r_m} \end{pmatrix} \quad \text{and} \quad W_m = \begin{pmatrix} w_1^\top \\ w_2^\top \\ \vdots \\ w_{r_m}^\top \end{pmatrix}, \quad (13)$$

that is, $v_i$ and $w_i^\top$ are the column vector of $V_m$ and row vector of $W_m$ respectively. Next, let $|v_i^{(\max)}|$ and $|w_i^{(\max)}|$ be the maximum absolute element of $v_i$ and $w_i$ respectively. Then, we define

$$D_V := \text{diag}(|v_1^{(\max)}|, |v_2^{(\max)}|, \ldots, |v_{r_m}^{(\max)}|) \quad (14)$$

and

$$D_W := \text{diag}(|w_1^{(\max)}|, |w_2^{(\max)}|, \ldots, |w_{r_m}^{(\max)}|). \quad (15)$$

Finally, we obtain

$$V_m' := V_m D_V^{-1}, \quad W_m' := D_W^{-1} W_m, \quad \text{and} \quad D_m := D_V D_W. \quad (16)$$

After the process described above, the maximum element of each column in $V_m'$ and that of each row in $W_m'$ is ±1, which resembles a kind of *scaling*. The additional factor, i.e. the diagonal matrix $D_m$, is helpful for avoiding overflow or underflow if the original $V_m$ or $W_m$ has an extremely large or small element. In addition, this part may also contribute to enlarge the precision (i.e. mantissa) because we can use the additional bits involving $D_m$. The required additional space is $O(r_m)$, which can be disregarded; the additional arithmetic cost is also $O(r_m)$, which is much smaller than those in the other parts of the low-rank submatrix vector multiplication.

Based on this extended data structure for low-rank submatrices in H-matrix, we then assign the following three combinations of precision:

- **Method 2-Double** (see Figure 3a):
  FP64 H-Matrix (Extended, with FP64 $D_m$) × FP64 Source Vector → FP64 Result Vector
- **Method 2-Single** (see Figure 3b):
  FP32 H-Matrix (Extended, with FP64 $D_m$) × FP32 Source Vector → FP64 Result Vector
- **Method 2-Mixed** (see Figure 3c):
  FP32 H-Matrix (Extended, with FP64 $D_m$) × FP64 Source Vector → FP64 Result Vector

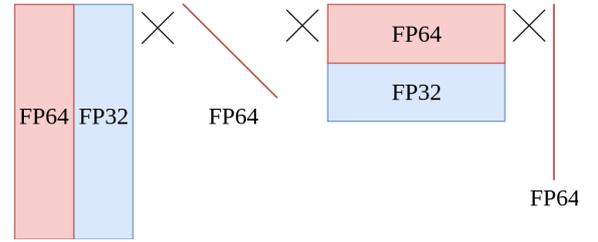

Figure 4: The low-rank matrix vector multiplication model for Method 3. Here we see that the submatrices are partially described in two different data types.

It is worth nothing that no matter which precision (FP64 or FP32) H-matrix is described in, the diagonal matrix $D_m$ is always being stored in FP64. The computation model can be seen in Figure 3.

### 3.3 Method 3: Mixed precision computing using a further modified data structure

Finally, for our third method, we further modify the data structure proposed in the second method in effort to enable more sophisticated combinations of precision types. Let us suppose an addition of two FP numbers, where one is relatively much larger than the other, e.g. $1.0 \times 10^0$ and $1.0 \times 10^{-20}$ in FP64. In this case, the latter is completely ignored in the operation. Motivated by this nature of IEEE 754's FP [1], we introduce a further modification into the data structure for representing each low-rank submatrix.

Since the middle diagonal matrix $D_m$ in our extended low-rank representation consists of the maximum elements (in terms of absolute value) of each column vector of $V_m$ and each row vector of $W_m$, from $D_m$, we can know the relative importance of each column in $V_m$ and row in $W_m$. Using this information, we split columns in $V_m$ (and rows in $W_m$) into two groups: columns (rows) stored in FP64 and those in FP32. Let

$$D_m = \text{diag}(d_1, d_2, \ldots, d_{r_m}), \quad (17)$$

and define

$$d_m^{(\max)} := \max\{d_1, d_2, \ldots, d_{r_m}\}. \quad (18)$$

Table 1: Summary of the precision types used in our proposed methods.

| Method | | H-matrix | | | | | | | | | Result Vector |
|---|---|---|---|---|---|---|---|---|---|---|---|
| | | Dense | | | Low-Rank | | | | | | |
| | | $\hat{y}_m$ | $A|^m$ | $x_m$ | $z_m$ | $W'_m$ | $x_m$ | $D_m$ | $\hat{y}_m$ | $V'_m$ | $\widetilde{y}$ |
| 1 | Double | FP64 | FP64 | FP64 | FP64 | FP64 | FP64 | – | FP64 | FP64 | FP64 |
| | Single | FP64 | FP32 | FP32 | FP32 | FP32 | FP32 | – | FP64 | FP32 | FP64 |
| | Mixed | FP64 | FP32 | FP64 | FP32 | FP32 | FP64 | – | FP64 | FP32 | FP64 |
| 2 | Double | FP64 | FP64 | FP64 | FP64 | FP64 | FP64 | FP64 | FP64 | FP64 | FP64 |
| | Single | FP64 | FP32 | FP32 | FP64 | FP32 | FP32 | FP64 | FP64 | FP32 | FP64 |
| | Mixed | FP64 | FP32 | FP64 | FP64 | FP32 | FP64 | FP64 | FP64 | FP32 | FP64 |
| 3 | | FP64 | FP64 | FP64 | Combination of Method 2 Double and Mixed | | | | | | FP64 |

Then, we introduce a constant integer $c$ and define

$$I_m^{(\text{FP32})} := \{i \mid d_i < d_m^{(\text{max})} \times 10^{-c}\} \quad (19)$$

and

$$I_m^{(\text{FP64})} := \{i \mid d_i \geq d_m^{(\text{max})} \times 10^{-c}\}. \quad (20)$$

Here, $c$ can be regarded as the *FP exponential precision distance* between $d_m^{(\text{max})}$ and $d_i$. Using the result of classification, namely $I_m^{(\text{FP32})}$ and $I_m^{(\text{FP64})}$, we store the $i$-th column of $V'_m$ and $i$-th row of $W'_m$ in FP32 if $i \in I_m^{(\text{FP32})}$, and those in FP64 if $i \in I_m^{(\text{FP64})}$. The diagonal matrix $D_m$ is always stored in FP64.

In this study, we consider the cases that $c = -1, 1, 2, \ldots, 7$. Observe that all of the dense submatrices are stored in FP64 in Method 3, and when $c = -1$, it means that all of $V'_m$ and $W'_m$ are stored in FP32. Figure 4 shows the computation model for low-rank submatrices in Method 3, where a submatrix have a part of itself in FP64 and another part of itself in FP32. As shown in the figure, we can reorder the columns of $V'_m$, diagonal elements in $D_m$, and rows of $W'_m$ without any loss of generality.

### 3.4 Summary of proposed methods

Here, we summarize our proposed methods. In order to clarify which precision (FP64 or FP32) is used for each data and calculation, we recall the essential computational steps. If a submatrix is dense, its (partial) matrix vector multiplication is calculated through the nested loops, and its core is

$$\hat{y}_m[i] = \hat{y}_m[i] + (A|^m[i][j] \times x_m[j]). \quad (21)$$

If a submatrix is low-rank, its (partial) matrix vector multiplication is calculated by two standard matrix vector multiplications (and an additional vector update in Method 2 and 3), and its essence is

$$z_m[i] = z_m[i] + (W'_m[i][j] \times x_m[j]), \quad (22)$$

$$z_m[i] = z_m[i] \times D_m[i][i], \quad (23)$$

which is not required in Method 1, and

$$\hat{y}_m[i] = \hat{y}_m[i] + (V'_m[i][j] \times z_m[j]), \quad (24)$$

where we replace $W'_m$ and $V'_m$ with $W_m$ and $V_m$ respectively in the case of Method 1. Finally, the Result Vector is obtained by

$$\widetilde{y}[i] = \widetilde{y}[i] + \hat{y}_m[i - i_m^{(s)} + 1]. \quad (25)$$

For Method 3, we separate the computation for each low-rank submatrix into Double and Mixed parts, employing the process in Method 2 Double and Mixed respectively, and then add the partial results ($\hat{y}_m$ from the Double part and that from the Mixed part).

Table 1 presents the precision types used in each method. Our program code for each method basically follows Equations (21) to (25). While it depends on the implementation of the machine, usually when precision types of two operands differ, the data with the lower precision is first cast into the higher precision before the operation is executed.

### 3.5 Krylov iterative solver using Mixed precision H-matrix vector multiplication

Since we evaluate the effectiveness of our proposed methods for mixed precision H-matrix vector multiplication in a Krylov iterative solver appearing in BEM analysis, we shall briefly explain one of the most popular iterative solvers for a linear system with a non-symmetric coefficient matrix solver, namely the BiConjugate Gradient STABilized (BiCGSTAB) method [26].

In this study, we employ the BiCGSTAB solver implemented in the conventional HACApK library and simply replace the original H-matrix vector multiplication with our proposed mixed precision H-matrix vector multiplication. The pseudocode of BiCGSTAB is provided as Algorithm 1, here we apply each of our proposed mixed precision H-matrix vector multiplication methods to the calculation of the product of $\widetilde{A}$ and a vector (i.e. $x_0, p_i, s$). In the case of when a Source Vector is described at FP32 in some of the methods, we first prepare a new FP32 vector variable, and then assign the original FP64 variable into our newly created FP32 variable just before each H-matrix vector multiplication. This means that the change of precision of the Source Vector does not effect other computations, e.g. vector inner product, in BiCGSTAB.

Generally, H-matrix vector multiplication is a memory-bound kernel, that is, the amount of memory access essentially limits the performance (i.e. execution time), which is one of the major bottlenecks of modern computers [8]. And as shown in Algorithm 1, BiCGSTAB heavily uses H-matrix vector multiplication, namely twice in each iteration, which the densely repeated H-matrix vector multiplications usually become the dominant part in execution time. In our proposed mixed precision computing methods, we reduce the amount of data for H-matrix by introducing FP32 data type. Thus, an acceleration of H-matrix vector multiplication is expected. And therefore, the accelerated mixed precision H-matrix vector multiplication naturally contributes to the reduction of execution time of each iteration in BiCGSTAB.

On the other hand, we need to pay attention to the drawbacks of using lower precision. Here, the number of iterations in BiCGSTAB strongly effects the execution time, and since the sacrifice of accuracy in mixed precision H-matrix vector multiplication may increase the number of required iterations, this approach may cause a worse convergence ratio. Hence, we evaluate our mixed precision methods for H-matrix vector multiplication not only on themselves but also in the context of iterative solvers. And so, we have to balance the trade-off between the speedup and the loss of accuracy in mind when using lower precision.

**Algorithm 1** BiCGSTAB with Mixed Precision H-Matrix vector multiplication

---

$x_0$ is an initial guess; $r_o = b - \widetilde{A}x_0$  // H-matvecmul occurs here
Choose $\widetilde{r}$, for example, $\hat{r} = r_0$
**for** $i = 1, 2, \ldots$ **do**
    $\rho_{i-1} = \widetilde{r}^\top r_{i-1}$
    **if** $\rho_{i-1} = 0$ **then**
        method fails
    **end if**
    **if** $i = 1$ **then**
        $p_i = r_{i-1}$
    **else**
        $B_{i-1} = (\rho_{i-1}/\rho_{i-2})(\alpha_{i-1}/\omega_{i-1})$
        $p_i = r_{i-1} + B_{i-1}(p_{i-1} - \omega_{i-1}v_{i-1}$
    **end if**
    $v_i = \widetilde{A}p_i$;  // H-matvecmul occurs here
    $\alpha_i = \rho_{i-1}/\widetilde{r}^\top v_i$
    $s = r_{i-1} - \alpha_i v_i$
    check $\|s\|_2$, if small enough: $x_i = x_{i-1} + \alpha_i p_i$ and stop
    $t = \widetilde{A}s$  // H-matvecmul occurs here
    $\omega_i = t^\top s/t^\top t$
    $x_i = x_{i-1} + \alpha_i p_i + \omega_i s$
    $r_i = s - \omega_i t$
    check convergence; continue if necessary
    or continuation it is necessary that $\omega_i \neq 0$
**end for**

## 4 EXPERIMENT AND RESULT

In this section, we report the results of our experiments, where we evaluate the effectiveness of our proposed mixed precision computing methods for H-matrix vector multiplication. We present the results of H-matrix vector multiplication itself and those in the context of the BiCGSTAB solver.

### 4.1 Experimental Settings

*4.1.1 Machine Setup.* For the experiments, we used Kyoto University's Laurel 2 Supercomputer (System B) to generate the empirical results, see Table 2. Each node has 2 Intel Xeon Broadwell processors running at a clock speed of 2.1 GHz. Each processor has 18 cores, totaling 36 cores per node. Our experiments are ran and evaluated from 1 thread up to 36 threads, fully utilizing 1 node.

**Table 2: Experiment setup and machine specification**

| Node | Processor (Core) | 2 (2 x 18 = 36) |
|---|---|---|
|  | Performance | 1.21 TFlops |
|  | Memory | 128GB |
|  | Injection Bandwidth | 12.1 GB/sec |
|  | Interconnect | Omni-path |
| Processor | Processor | Intel Xeon Broadwell |
|  | Architecture | x86-64 |
|  | Clock | 2.1 GHz |
|  | Number of Cores | 18 |
|  | Performance | 605 GFlops |
| Software | Programming Language | Fortran 90 |
|  | H-Matrix Libary | HACApK |
|  | Multithreading Library | OpenMP |
|  | BLAS Library | Intel MKL |
|  | Compiler | Intel Compiler v17.0.2 |
|  |  | -qopenmp -O3 -ip |

*4.1.2 Test Problem: Surface Charge Analysis.* For the problem of our experiments, we test on an electrostatic field analysis that is solved using BiCGSTAB. The goal of this analysis is to compute the surface charge densities that are generated when multiple perfectly conductive spheres in a 3-dimensional space are excited. The problem is adopted from a previous work using H-Matrix for parallel BEM analyses [17].

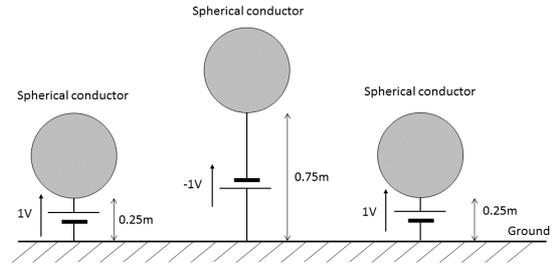

**Figure 5: Model of surface charge analysis.**

The summary is that 3 perfectly conductive spheres are set at a fixed distance away from each other and excited with a fixed voltage which is shown in Figure 5. We make use of Method of Images (MoI), resulting in 6 spheres, consisting 3 originals and 3 images. The induced electric surface charge is then computed with BEM which the 6 spheres are discretized with 64,800 triangular face elements. In other words, there are 64,800 unknowns to solve for this linear system. The dense coefficient matrix is first approximated to become a H-Matrix so that each submatrix can be processed in parallel. Then, the linear system whose coefficient matrix is given as H-matrix is solved by an iterative solver.

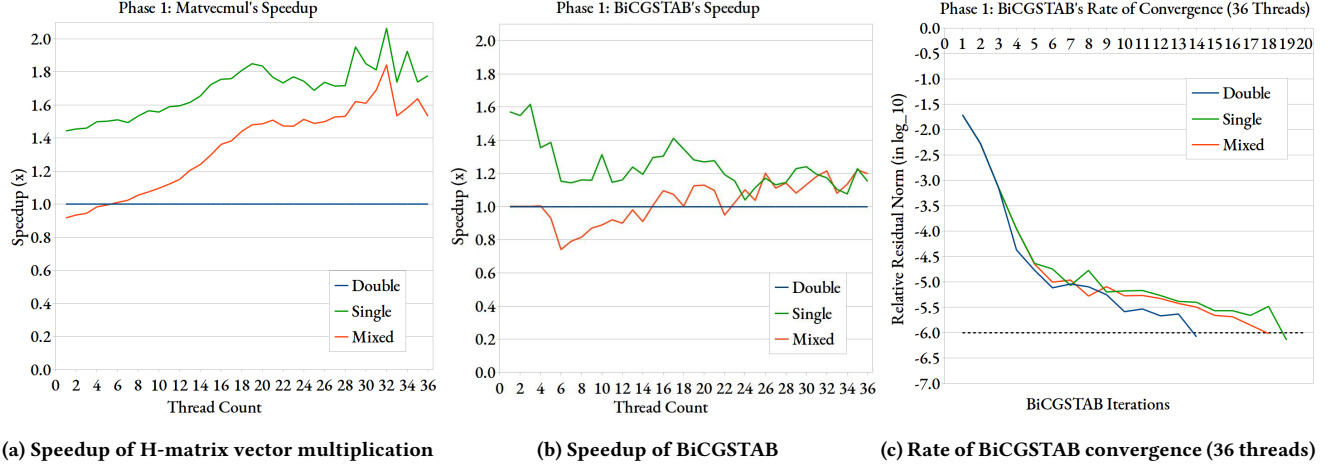

(a) Speedup of H-matrix vector multiplication　　(b) Speedup of BiCGSTAB　　(c) Rate of BiCGSTAB convergence (36 threads)

**Figure 6: Summarized experiment results for Method 1.**

*4.1.3 Other Settings.* The coefficient H-matrix is first generated in FP64 by the process provided in the HACApK library. Then, each submatrix is converted into FP32 if required by our proposed methods. In Method 2 and 3, we first decompose the low-rank submatrices in FP64, resulting in the generation of $D_m$, and then apply the required conversion into FP32.

We set the convergence criterion to be $10^{-6}$; we determine the convergence if the relative residual norm $\|r_i\|_2/\|b\|_2$ is less than $10^{-6}$ in BiCGSTAB. It is worth noting that the computed $r_i$ in BiCGSTAB is actually affected by lower precision. Therefore, we compute the true residual using the original FP64 H-matrix and conclude that the solver converged only if the true relative residual norm is actually smaller than $10^{-6}$.

We do 10 sets of experiments for all experiments and average the computed execution time in seconds. For the pure evaluations of H-matrix vector multiplication without involving BiCGSTAB, each set of H-matrix vector multiplication is repeated 1000 times on the same source vector to obtain stabler results.

## 4.2 Result of Method 1

For H-matrix vector multiplication, Figure 6a shows the speedup over the baseline, i.e. Method 1-Double. Ideally, the speedup of FP32 over FP64 should be about 2.0 because the amount of transferred data memory access is approximately halved. Here, we observe that Method 1-Single is always faster than Method 1-Double, and the speedup approaches 2x with higher thread count. Now, Method 1-Mixed is slower than Method 1-Double at lower thread count initially but approaches Method 1-Single at higher thread count, that is, as the thread count increases, the speedup is generally better. This can be due to the fact that when at lower thread count, the memory bandwidth is effectively fully utilized where at higher thread count, the memory bandwidth becomes the major bottleneck and so we see a better speedup.

Additionally, note that the speedup seems to increase to 18 thread count, then dropped a bit, and kept increasing until it approaches 2x again. This can be explained by the fact that we are using 1 node that contains 2 processors, each 18 cores, essentially a thread affinity problem with higher cache miss probability. And interestingly, at 34 threads, Method 1-Single achieves 2.06x super speedup. This might be due to load balancing problem because H-Matrices contain submatrices that vary widely in sizes and it seems to be just a coincidence that 34 threads being the best performing parameter for the tested problem. The final result for Method 1-Double, Method 1-Single, and Method 1-Mixed at 36 threads are 13.757s, 7.743s, and 8.981s respectively, which Method 1-Single is the fastest.

For BiCGSTAB, looking at the speedup of BiCGSTAB in Figure 6b, where we compare the elapsed time for the BiCGSTAB to converge using different precision, we observe that the speedup by switching to FP32 is lesser with higher thread count. Method 1-Mixed again show the phenomenon of being initially slower than Method 1-Double at lower thread count but approaches Method 1-Single with higher thread count. This can be due to numerous reasons, but is most probably related to the hardware implementation of mixed precision computing. When multiplying two value of different data types, unless it is implemented to do mixed precision directly, usually either one of the values must be converted or cast to fit the other data type so that the multiplication can go through. At lower threads, the casting overhead might be so noticeable that they become even slower than the pure FP64 counterparts. At higher threads, these castings are parallelized and the overhead is less significant.

The final speedups of Method 1-Single and Mixed compared against the Double baseline are 1.152x and 1.198x respectively. This result shows that mixed precision computing works because non-FP64 versions are about 15% to 20% faster than the baseline FP64. The final result for Double, Single, and Mixed at 36 threads are 0.563s, 0.489s, and 0.470s respectively, which Mixed is the fastest.

Finally, Figure 6c definitely shows that the linear solver successfully converged even at lower precision, albeit using more iterations. Double, with the highest precision, requires the least iterations while Single, with the lowest precision, requires the most iterations. Mixed, with a precision in the middle, requires an iteration count

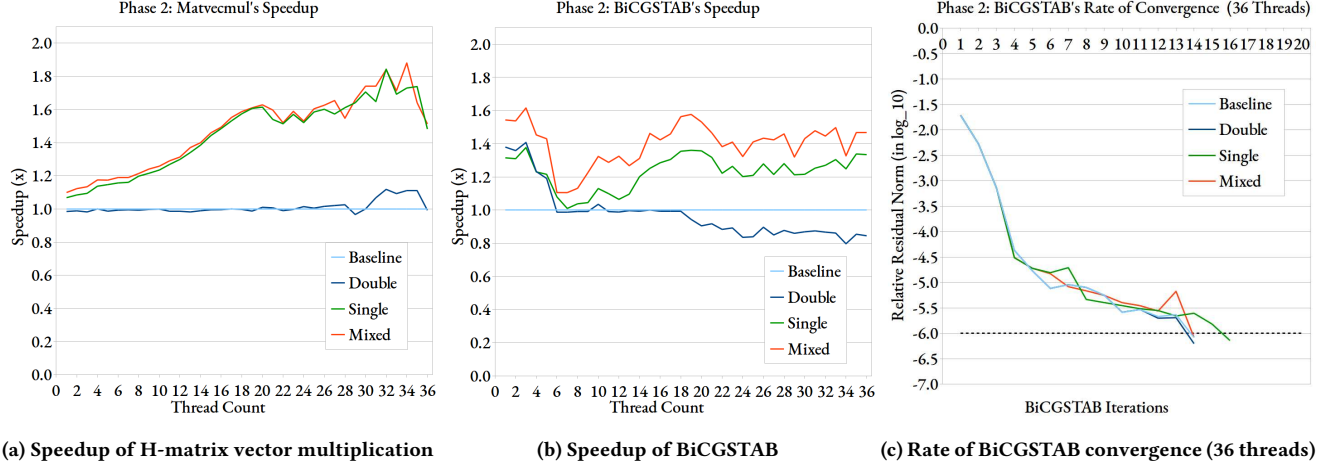

(a) Speedup of H-matrix vector multiplication  (b) Speedup of BiCGSTAB  (c) Rate of BiCGSTAB convergence (36 threads)

Figure 7: Summarized experiment results results for Method 2.

between Double and Single. Interestingly, unlike Single, the rate of convergence of Mixed did not halve (i.e. number of iterations required to converge did not double) and result in a faster elapsed real time. This implies that the required precision to solve the problem is actually between FP32 and FP64. That is, storing the matrix in FP32 means insufficient precision while storing it in FP64 means being space inefficient.

### 4.3 Result of Method 2

For H-matrix vector multiplication, differ from Method 1, here in Figure 7a, we show the speedup over Method 1-Double (Baseline), we observe that Method 2-Single and Method 2-Mixed are always faster than Method 2-Double. Both Method 2-Single and Method 2-Mixed attempt to approach ideal 2x speedup at the beginning but degrades at the very end. More threads are necessary to conclude if the speedup is approaching 2x with more threads or is starting to degrade from there due to overhead or other side effects. Also, we see that the performance of Method 2-Double perform similarly to Method 1-Double (Baseline). And, the speedup of Method 2-Single and Method 2-Mixed compared to Baseline at 36 threads are 1.481x and 1.513x respectively. The fastest speedup achieved by Method 2-Single and Method 2-Mixed compared to Baseline are 1.843x at 32 threads and 1.881x at 34 threads respectively. The final result for Method 2-Double, Method 2-Single, and Method 2-Mixed at 36 threads are 13.868s, 9.284s, and 9.094s respectively, which Method 2-Mixed is the fastest.

For BiCGSTAB, we also show the speedup over the case of using Method 1-Double (Baseline) in Figure 7b. Although Method 2-Double's matrix vector multiplication perform similarly to Baseline, Method 2-Double perform worse than Baseline. Method 2-Single and Method 2-Mixed are, however, always much faster than the Baseline, with Mixed having a clear lead. The speedup of Method 2-Single and Method 2-Mixed compared to Baseline at 36 threads are 1.334x and 1.467x respectively. The final result of Method 2-Double, Method 2-Single, and Method 2-Mixed are 0.666s, 0.422s, and 0.384s at 36 threads respectively.

Finally, as seen from Figure 7c, both Method 2-Double and Method 2-Mixed require only 14 steps to converge, similar to the Baseline. More importantly, even if the result factor in the extra computations needed due to the introduction of a diagonal part to the low rank approximation for both Method 2-Single and Method 2-Mixed versions, the number of iterations required to converge actually reduced instead of the opposite. This experiment definitely proved the reasonability of our idea described in Equation (11).

### 4.4 Result of Method 3

For H-matrix vector multiplication, Figure 8a shows the speedup of Method 3 with each $c$ setting over the Baseline (i.e. Method 1-Double). Here, when $c = -1$, the low-rank submatrices are wholly FP32, and achieves performance of 10.675s at 36 threads. Interestingly, from Figure 8a, we see a "flipping" behavior where the performance of those with higher $c$ are faster than those with lower $c$ at lower thread count initially but reverse and become slower at higher thread count. Note that there is another major flipping behavior occurring at $c = 1$ at lower threads. From $c = -1$ increasing to $c = 1$ at lower threads, the H-matrix vector multiplication loses speed; yet from $c = 1$ to $c = 7$ at higher threads, the H-matrix vector multiplication gains speed. Overall, the matrix vector multiplication is the faster when $c$ values are lower at higher threads.

For BiCGSTAB, Figure 8b shows the speedup over the case of using Method 1-Double (Baseline). As the value of $c$ decreases from 7 to 1, the performance degrades. However, when $c$ reaches 0, the BiCGSTAB performance flips around, increases speed dramatically, and surpasses Baseline. The performance is the fastest when $c = -1$. The final result when $c = -1$ at 36 threads is 0.437s.

Finally, Figure 8c shows the summarized rate of convergence of different parameters. They all successfully converge, but $c = 1$ and $c = 2$ require more than 20 steps and so are not shown in the figure. The mentioned flipping behavior is again seen in this figure. Now, we are taken aback by the results, especially when split criterion $c = -1$, the version with the pure FP32 low-rank submatrices and pure FP64 dense matrix, achieves the best result.

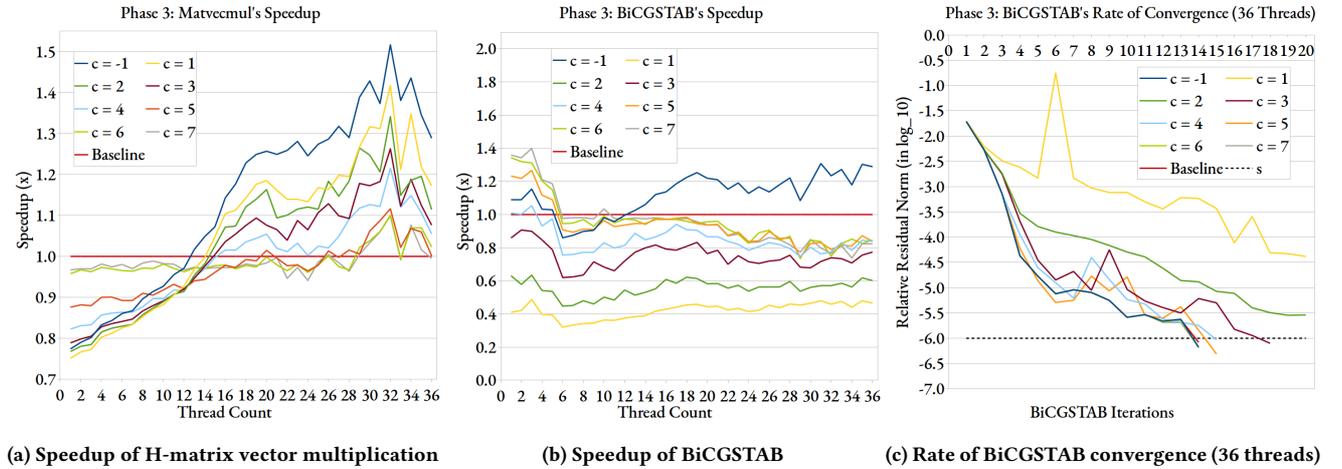

(a) Speedup of H-matrix vector multiplication    (b) Speedup of BiCGSTAB    (c) Rate of BiCGSTAB convergence (36 threads)

Figure 8: Summarized experiment results results for Method 3.

This implies that the dense FP64 submatrices almost or completely retain all precision as the BiCGSTAB convergence performs almost exactly the same as Baseline. This is despite we halved the precision of the low-rank submatrices from FP64 to FP32 and yet we do not lose any noticeable precision for the final result. This major insight leads us to believe that we can combine both ideas of Method 2 and Method 3 together for our future work and reinforces the theory that the required precision is actually between what is produced by Method 3's H-Matrix (FP64 Dense & FP32 Low-Ranks) and Method 2's H-Matrix (FP32 Dense & FP32 Low-Ranks).

## 4.5 Future Work

Moving forward from Method 3, we realize that the dense submatrices have probably retained most of the required precision to converge successfully. Hence, our attempt of lowering just the precision of the low-rank submatrices in Method 3 have had negligible impact on the accuracy of final result. We predict that, by lowering the precision of the dense submatrices instead, especially by splitting them into multi-level precision like Method 3, we would be able to hit the optimal precision combinations to best find the sweet spot of the precision-accuracy trade-off relationship.

In addition, if we were to go deeper by introducing additional precision, especially FP16, we open the door to utilizing NVIDIA's GPU Tensor Cores [13] for problems like deep learning where half precision may be sufficient. On the other hand, if a number is relatively much smaller than the maximum element, instead of going lower precision, we can also simply delete the element with its corresponding multiplicand rows and multiplier columns.

Indirectly related, the load balancing problem due to the the fact that the H-Matrix submatrices vary widely in sizes might also be a potential future work. This is particularly useful as if we have better knowledge of the load balance then we can have cleverer choices for determining which parts to describe at lower precision.

Finally, we would also like to transfer this contribution to other kinds of data structures and linear solving methods as we believe our study can be widely applied.

## 5 SUMMARY & CONCLUSION

In this paper, we propose three methods of mixed precision hierarchical matrix (H-matrix) computing. Concretely, we used three types of data structures, including two newly proposed ones for low-rank representation, and assigned them data types of different precision. We evaluated the performance of H-matraix vector multiplication and the BiCGSTAB solver, which utilizes heavy repetitions H-matrix vector multiplication. Ultimately, Method 2-Mixed attains ~1.5x speedup in BiCGSTAB over the baseline, namely Method 1-Double, where it not only accelerated the H-matrix vector multiplication portion, it also retains almost the same BiCGSTAB convergence rate as that of the baseline.

Our methods extend the HACApK, a H-matrix library, with mixed precision computing capability, focusing on the intensive H-matrix vector multiplication part. By employing the three proposed techniques described in the paper, such as describing parts of the H-matrix in lower precision, we gain speed as those parts are less important and do not significantly affect the rate of convergence or final result. And because parts of the H-matrix is in lower precision, we achieve better storage space and matrix vector multiplication time performance due to less data transfer and computation cost.

Furthermore, we achieved the goal of the paper not only by utilizing mixed precision computation between two data types, we also introduced a new kind of data structure where a submatrix can partially be both FP64 and FP32 (see Figure 4). By dividing data structure into multiple precision levels, the precision-performance trade-off can be intricately balanced so that the coefficient H-matrix is at its most efficient precision level, resulting in the most optimized space and computation performance. We intend to further deepen this multi-level precision data structure with more levels of precision, especially with FP16 or even FP8.

Finally, our study is potentially universal, and can be translated to numerous other solving techniques, including the Generalized Minimal RESidual (GMRES) method linear solver. This comes when the exploded interest in deep learning in recent years and approximating neural network model with H-matrix is taking foot [9, 10].

Due to the end of exponentially increasing FLOPS era [20], it is stimulating to see mixed precision computing playing a larger role on general computation and changing the dynamics of data-oriented computation. With so many next questions in mind and a clear path forward, we have an exciting future outlook on accelerating matrix vector multiplication, matrix matrix multiplication, and linear solvers with mixed precision H-matrix computation and data structure.

## ACKNOWLEDGMENTS

This work was supported by "Japan Society for the Promotion of Science" (JSPS) KAKENHI Grant Numbers JP19H04122 and JP17H01749. This work was supported by "Joint Usage/Research Center for Interdisciplinary Large-scale Information Infrastructures" (JHPCN) and "High Performance Computing Infrastructure" (HPCI) in Japan (Project ID: jh190042-NAH). Finally, we express our appreciation to Professor Yasuhito Takahashi and Professor Takeshi Mifune for their contributions in providing the BEM test model.